\documentclass[12pt,a4paper]{article}
\usepackage{amsmath}
\usepackage[font={footnotesize,it}]{caption}
\usepackage[top=1.2in, bottom=1.2in, left=1.1in, right=1.1in]{geometry}

\DeclareCaptionStyle{italic}[justification=centering]{labelfont={bf},textfont={it},labelsep=colon}
\usepackage{graphicx,psfrag,epsf}
\usepackage{enumerate}
\usepackage{natbib}
\usepackage{url, setspace,multirow}
\usepackage{booktabs, subfig, bm, paralist,mathpazo,tikz,todonotes,longtable,microtype,dsfont,rotating} 
\usepackage[pdftex,colorlinks=true]{hyperref}
\definecolor{darkblue}{rgb}{0,0,.6}
\hypersetup{citecolor=darkblue,linkcolor=darkblue,urlcolor=darkblue}

\newcommand{\blind}{0}

\graphicspath{{plots/}}

\newsavebox\CBox

\date{\today}

\begin{document}

\def\spacingset#1{\renewcommand{\baselinestretch}
{#1}\small\normalsize} \spacingset{1}

\if0\blind
{
  \title{\bf Forecasting functional time series using weighted likelihood methodology}
  \author{
  Ufuk Beyaztas \\
  Department of Statistics \\
  Bartin University \\
  \\
  Han Lin Shang\\
    Research School of Finance, Actuarial Studies and Statistics \\
    Australian National University}
  \maketitle
} \fi

\if1\blind
{
   \title{\bf Forecasting functional time series using weighted likelihood methodology}
   \author{}
   \maketitle
} \fi

\bigskip

\begin{abstract}
Functional time series whose sample elements are recorded sequentially over time are frequently encountered with increasing technology. Recent studies have shown that analyzing and forecasting of functional time series can be performed easily using functional principal component analysis and existing univariate/multivariate time series models. However, the forecasting performance of such functional time series models may be affected by the presence of outlying observations which are very common in many scientific fields. Outliers may distort the functional time series model structure, and thus, the underlying model may produce high forecast errors. We introduce a robust forecasting technique based on weighted likelihood methodology to obtain point and interval forecasts in functional time series in the presence of outliers. The finite sample performance of the proposed method is illustrated by Monte Carlo simulations and four real-data examples. Numerical results reveal that the proposed method exhibits superior performance compared with the existing method(s).
\\

\end{abstract}

\newpage
\spacingset{1.48}

\section{Introduction\label{sec:intro}}

A functional time series arises when functional objects are collected sequentially over time. In other words, let $\left\lbrace y_t \right\rbrace $ ($t = 1, \cdots, N$) denote a sequence of observed functional data; then, it is termed an functional time series if each functional element $y_t(\tau)$ is defined on a bounded interval with a continuous variable $\tau$; $\tau \in [a,b]$. Denote by $\left\lbrace y_t(x_j) \right\rbrace$, for $t = 1, \cdots, N$ and $j = 1, \cdots, J$, the noisy functional time series data observed at time points $\left\lbrace x_1, \cdots, x_J\right\rbrace$. It is assumed that the functional time series is characterized by a continuous smooth function $f$ and an error process $\varepsilon$ such that
\begin{equation} \label{eq:gen}
y_t(x_j) = f_t(x_j) + \sigma_t(x_j) \varepsilon_{t,j}
\end{equation}
where $\varepsilon_{t,j}$ is an independently and identically distributed (i.i.d.) Gaussian random variable with zero mean and unit variance and $\sigma_t(x_j)$ allowing for heteroskedasticity.

Forecasting unobservable future realizations of functional time series is of great interest. In practice, forecasts can be obtained in the form of point and/or interval forecasts. A point forecast corresponds to an estimate (conditionally on the available data) of the unknown future realization of the underlying process. However, this approach may not produce reliable inferences for future observations since it does not provide any information about the degree of uncertainty associated with the point forecasts. By contrast, interval forecasts, such as the prediction interval, provide better inferences taking into account the uncertainty  associated with point forecasts; see for example \cite{Chatfield1993}, \cite{Kim2001}, and \cite{JoreMitchellVahey2010}. In the context of functional time series, \cite{HydUl} propose a functional data approach to obtain point and interval forecasts for age-specific mortality and fertility rates observed over time. While doing so, they consider that the functional data $\left\lbrace y_t(x_j) \right\rbrace$ consist of random functions separated by consecutive and continuous time intervals. Their approach is as follows: 
\begin{inparaenum}
\item[($i$)] Approximate the smooth functions $\left\lbrace f_t(x) \right\rbrace$ separately using a nonparametric smoothing technique on each function $y_t(x)$ ($t = 1, \cdots, N$). 
\item[($ii$)] Decompose the smoothed functional time series data into $K$ orthonormal principal components $\phi_k$ ($k = 1, \cdots, K$) and associated uncorrelated scores $\beta_{k,t}$ ($t = 1, \cdots, N$) using a functional principal component model. 
\item[($iii$)] Apply a univariate time series model to each principal component score to obtain their future values.
\item[($iv$)] Calculate the future realization of the functional time series by multiplying principal components with the forecasted part of the principal component scores.
\item[($v$)] Obtain the prediction intervals under the assumption of Gaussian distributed error terms. This approach (or a modified version) has received extensive attention in the literature and has successfully been used in many areas; see for example, \cite{HydSh2009}, \cite{ShHyd2011}, \cite{Shang2013}, \cite{Kosiorowski2014}, \cite{GaoSh}, \cite{WMuns}, \cite{curceac} and references therein.
\end{inparaenum}

The aforementioned literature uses non-robust time series estimation methodologies when forecasting future values of the principal component scores. However, this approach is affected by the outliers which are common in real data sets; see, for example, \cite{Shang2019R}. An outlier is an observation that has been generated by a stochastic process with a distribution different from that of the vast majority of the remaining observations \citep{Rana}. In the context of functional time series, three types of outliers are observed: 
\begin{inparaenum}
\item[(1)] \textit{magnitude outlier}, which is a point far from the bulk of the data; 
\item[(2)] \textit{shape outlier}, which falls within the range of the data but differs in shape from the bulk of the data; and 
\item[(3)] the combination of both outlier types. 
\end{inparaenum}
For more information about the outliers in functional time series, see \cite{Febreroetal}, \cite{HS2010}, \cite{sun} and \cite{Rana}. 

In the case of outlying observation(s), non-robust techniques produce biased estimates, and high forecasting errors correspond to outlying observations. In such cases, the high forecasting errors may severely affect point forecasts as well as prediction intervals and could lead to unreliable inferences. We propose a robust functional time series forecasting method based on the minimum density power divergence estimator of \cite{Basu1998}. The proposed method uses a modified version of the usual maximum likelihood score equations, called weighted score equations, to estimate the model parameters. The weighted score equations are defined as a function of the Pearson residuals for which large values are obtained when the observations diverge from the underlying model. Hence, this approach makes it possible to check whether the maximum likelihood estimators are affected by a set of observations that are inconsistent with the model, and provides robust estimates by downweighting such observations. It also provides weighted residuals, which are used to obtain point and/or interval forecasts. We use several Monte Carlo experiments and real-data examples to compare the finite sample performance of the proposed and existing methods. Our numerical records, which are discussed in Sections~\ref{sec:sim} and~\ref{sec:real}, reveal that the proposed method provides finite sample performance competitive with that of the existing methods when no outlier is present in the observed data. In addition, when outliers are present in the data, its performance is shown to be superior compared with that of available techniques.

The remaining of this paper is organized as follows. Section~\ref{sec:meth} provides an overview of the functional time series methods considered and the weighted likelihood estimation methodology. Several Monte Carlo experiments under different scenarios are conducted to evaluate the finite sample performance of the proposed method, and the results are presented in Section~\ref{sec:sim}. Section~\ref{sec:real} reports the findings obtained by applying the proposed method to some real-data examples. Section~\ref{sec:conc} concludes the paper.

\section{Methodology\label{sec:meth}}

Let us consider a sequence of stationary functional time series $\left\lbrace y_t(x): t \in \mathcal{Z}, x \in J \right\rbrace$ where $J$ is a bounded interval. It is assumed that the functions $y_t$ are elements of the metric, semi-metric, Hilbert or the Banach space, in general. We assume that the functions are elements of a square-integrable function $y \in L^2[0,1]$ residing in Hilbert space $\mathcal{H}$ satisfying $\int y^2(x)dx < \infty$ with an inner product $\langle y,z \rangle = \int y(x) z(x) dx$, $\forall y, z \in L^2[0,1]$. Denote by $\left( \Omega, \Sigma, P\right)$ the probability space where $\Omega$, $\Sigma$ and $P$ represent the sample space, $\sigma$-algebra on $\Omega$ and the probability measure on $\Sigma$, respectively. Then the random functional variable $y$ is defined as $y: \left( \Omega, \Sigma, P\right) \rightarrow \mathcal{H}$ so that $y$ is assumed to be an element of $L^2$ and $y^{-1}\left( \mathcal{B}\right) \in \Sigma$ where $\mathcal{B}$ is a Borel set of the Borel $\sigma$-algebra generated by $L^2$. We further assume that the random variable $y \in L^2 \left(\Omega\right)$ with finite second-order moment is a second-order stochastic process so that $E \left[ \vert y \vert^2 \right] = \int_{\Omega} \vert y \vert^2 dP < \infty$. The mean and covariance functions of the random variable $y$ are defined as in (\ref{eq:mean}) and (\ref{eq:cov}), respectively; see \cite{RamDal1991}.
\begin{eqnarray}
\mu(x) &=& E \left[ y(x) \right] = \int_{\Omega} y(x) dP \label{eq:mean} \\
C(x,s) &=& E \left[ \left( y(x) - \mu(x) \right) \left( y(s) - \mu(s) \right) \right] \nonumber \\
&=& \int_{\Omega} \left[ \left( y(x) - \mu(x) \right) \left( y(s) - \mu(s) \right) \right] dP \label{eq:cov}
\end{eqnarray}
Let $\left\lbrace y_t(x): t = 1, \cdots, N, x \in J \right\rbrace$ be an observed functional time series of size $N$ with the same distribution as $y$. Then, the sample mean and sample covariance functions are given by (\ref{eq:smean}) and (\ref{eq:scov}), respectively.
\begin{eqnarray}
\bar{\mu}(x) &=& N^{-1} \sum_{t=1}^N y_t(x) \label{eq:smean} \\
\hat{C}(x,s) &=& (N-1)^{-1} \sum_{t=1}^N \left( y_t(x) - \bar{y}(x) \right) \left( y_t(s) - \bar{y}(s) \right) \label{eq:scov}
\end{eqnarray}

Functional principal component analysis is frequently used to analyse functional time series. Briefly, it represents the data by a linear combination of orthonormal principal components $\phi_k$ and their associated scores $\beta_{k,t}$ ($k = 1, 2, \cdots$). In doing so, it decomposes the covariance operator given in (\ref{eq:cov}) into orthogonal bases of eigenfunctions. Let $\psi_k$ and $\lambda_k$, respectively, denote the $k$\textsuperscript{th} eigenfunction and eigenvalue. Then, the covariance operator is decomposed as follows:
\begin{equation*}
C(x,s) = \sum_{k=1}^{\infty} \lambda_k \psi_k(x) \psi_k(s).
\end{equation*}
The $k$\textsuperscript{th} principal component score is then defined as $\beta_{k,t} = \int y_t(x) \psi_k(x) dx$. In what follows, the random functions are expressed using Karhunen-Lo\`{e}ve expansion as:
\begin{equation*}
y_t(x) = \sum_{k=1}^{\infty} \beta_{k,t} \psi_k(x)
\end{equation*}
See \cite{RamSil2002}, \cite{Ramsayetall2009} and \cite{Shang2014} for more details about functional principal component analysis and its practical demonstration.

Let $\left\lbrace y_t(x_j) \right\rbrace$, for $t = 1, \cdots, N$ and $j = 1, \cdots, J$ be the observed functional time series (with noise) characterized by a continuous smooth function $f$ and an error process $\varepsilon$:
\begin{equation*}
y_t(x_j) = f_t(x_j) + \sigma_t(x_j) \varepsilon_{t,j}
\end{equation*}
Each function can be decomposed using a basis function expansion, as follows:
\begin{equation}\label{eq:fpca}
f_t(x) = \hat{\mu}(x) + \sum_{k=1}^K \hat{\phi}_k(x) \hat{\beta}_{k,t} + \epsilon_t(x) 
\end{equation}
where $\hat{\mu}(x) = N^{-1} \sum_{t=1}^N f_t(x)$ is the estimated mean of functions, $\hat{\phi}_k(x)$ and $\hat{\beta}_{k,t}$ denote the estimated functional principal components and their associated principal component scores, respectively, $\epsilon_t(x)$ is the error function with mean zero and $K$ is the number of basis functions with $K < N$. The choice of the basis function in (\ref{eq:fpca}) is arbitrary under orthogonality restriction. As \cite{HydUl} highlight, each principal component score can be forecasted independently using univariate time series methods since they are uncorrelated to each other. They also note that there may be cross-correlations at non-zero lags but these can be negligible. However, multivariate time series methods, such as vector autoregression, can be used to take into account the cross-correlations; see \cite{Aue}. Let $h$ represent the forecast horizon. Denote by $\hat{\beta}_{k, N+h|N}$ the $h$-step-ahead forecast of $\beta_{k, N+h}$ conditionally on the available data up to time $N$. Then, the point predictor of $y_{N+h}(x)$ is obtained by multiplying the forecasted principal component scores with the estimated functional principal components, as follows:
\begin{equation*}
\hat{y}_{N+h|N} = \hat{\mu}(x) + \sum_{k=1}^K \hat{\phi}_k(x) \hat{\beta}_{k, N+h|N}
\end{equation*}

As noted in Section~\ref{sec:intro}, point forecasts do not provide any information about the uncertainty of future realization of the functional time series. Conversely, the prediction interval is capable of producing valid inferences taking into account the uncertainty of each forecast. \cite{HydUl} propose a $(1-\alpha)100\%$ prediction interval for $y_{N+h}$ under the assumption of normality. However, this approach may seriously be affected by any departure from the normality assumption, which is not known in practice. In such cases, the bootstrap method is commonly used method to overcome this issue since it does not require full knowledge of the underlying distributional assumption. Therefore, we consider only the bootstrap prediction interval. \cite{HydSh2009} introduce a bootstrap approach to construct prediction intervals for the future values of the functional time series. Based on the functional time series model discussed above, we incorporate three error sources to construct the bootstrap prediction interval: 
\begin{inparaenum}
\item[($i$)] smoothing error $\varepsilon_{t,j}$ given in (\ref{eq:gen}), 
\item[($ii$)] the error caused by the functional principal component decomposition $\epsilon_t(x)$ in (\ref{eq:fpca}), and ($iii$) the error occurring owing to forecasting principal component scores $\beta_{k,t}$. Let $\xi_{k,h,t}$ denote the $h$-step-ahead forecast error $\hat{\xi}_{k,h,t} = \hat{\beta}_{k,t} - \hat{\beta}_{k, t|t-h}$ ($t = h+1, \cdots, N$). Then, the algorithm of the bootstrap procedure proposed by \cite{HydSh2009} is as follows.
\end{inparaenum}
\begin{itemize}
\item[Step 1.] Smooth the entire functional time series $y_t(x) = f_t(x) + \sigma_t(x) \varepsilon_{t}$ for $t = 1, \cdots, N$ to obtain smooth functions $f_t(x)$, smoothing error $\hat{\varepsilon}_{t}$ and variance component $\hat{\sigma}_t(x)$.
\item[Step 2.] Decompose the smooth functions into $k = 1, \cdots, K$ orthonormal functional principal components and associated principal component scores and obtain the fitted functions $\hat{f}_t(x)$. Then, calculate the error functions $\hat{\epsilon}_t(x) = f_t(x) - \hat{f}_t(x)$.
\item[Step 3.] Obtain $h$-step-ahead forecasts of the principal component scores $\beta_{k,t}$, $\hat{\beta}_{k, N+h|N}$ for $k = 1, \cdots, K$ and $t = N+1, \cdots, N+h$ using a univariate time series method. In addition, obtain the $h$-step-ahead forecast errors $\hat{\xi}_{k,h,t}$.
\item[Step 4.] Calculate the future bootstrap values as follows:
\begin{equation*}
\hat{y}^*_{N+h|N}(x) = \hat{\mu}(x) + \sum_{k=1}^K \hat{\beta}^*_{k, N+h|N}  + \hat{\epsilon}^*_{N+h|N}(x) + \hat{\sigma}^*_{N+h}(x) \hat{\varepsilon}^*_{N+h|N,j}
\end{equation*}
where $\hat{\beta}^*_{k, N+h|N} = \hat{\beta}_{k, N+h|N} + \xi_{k,h,*}$, $\xi_{k,h,*}$ is an i.i.d. random sample from $\left\lbrace \xi_{k,h,t} \right\rbrace$, $\hat{\epsilon}^*_{N+h|N}(x)$, $\hat{\varepsilon}^*_{N+h|N,j}$ and $\hat{\sigma}^*_{N+h}(x)$ are random samples with replacement from $\left\lbrace \hat{\epsilon}_t(x) \right\rbrace$, $ \left\lbrace \hat{\varepsilon}_{t} \right\rbrace$ and $\left\lbrace \hat{\sigma}_t(x) \right\rbrace$,  respectively. 
\item[Step 5.] Repeat Step 4. $B$ times by drawing random samples of the error terms to obtain $B$ sets of bootstrap replicates of $\hat{y}^*_{N+h|N}(x)$, $\left\lbrace \hat{y}^{*,1}_{N+h|N}(x), \cdots, \hat{y}^{*,B}_{N+h|N}(x) \right\rbrace$ for each $h$, where $B$ denotes the number of bootstrap simulations.
\end{itemize}
The $(1-\alpha)100\%$ bootstrap prediction intervals for $y_{N+h}$ are then obtained by the $\alpha/2$\textsuperscript{th} and $(1-\alpha/2)$\textsuperscript{th} quantiles of the  bootstrap replicates $\left\lbrace \hat{y}^{*,1}_{N+h|N}(x), \cdots, \hat{y}^{*,B}_{N+h|N}(x) \right\rbrace$. This approach works well when the functional time series is relatively smooth, and no outlier is present in the data. It has successfully been used in a wide range of applications; see, for example, \cite{Hdem}, \cite{Anerios}, \cite{Husin}, \cite{Anerios2016}, \cite{Canale}, \cite{Shangsp500}, \cite{Kearney} and \cite{ShangAOR}. By contrast, the traditional univariate time series methods used for modelling the functional principal component scores may produce biased estimates as well as high forecasting errors, leading to unreliable results when outliers are present in the data. To overcome this issue, we propose a robust forecasting functional time series method. It is based on replacing the traditional estimators and residuals by the minimum power divergence estimator of \cite{Basu1998} and weighted residuals, respectively.

\subsection{The weighted likelihood methodology}

To employ the weighted likelihood methodology in the forecasting of functional time series, we consider the stationary autoregressive model of order $p$ (AR($p$)). Let us suppose that the $k$\textsuperscript{th} principal component score $\beta_{k,t}$ ($t=1,\cdots,N$) is characterized by a zero mean AR($p$) process, as follows:
\begin{equation} \label{eq:ar}
\beta_{k,t} = \phi_1 \beta_{t-1,k} + \cdots, \phi_p \beta_{t-p,k} + \zeta_t, ~~ t=1,\cdots,N
\end{equation}
where $\Phi=(\phi_1, \cdots, \phi_p)$ denotes the parameter vector and $\zeta_t$ is an i.i.d. white noise sequence with mean zero and variance $\sigma^2_{\zeta}$. Under the assumption of normality, the probability density function for the model (\ref{eq:ar}) is given by
\begin{equation} \label{eq:dens}
P(\zeta_N|\Phi;\sigma^2_{\zeta}) = \left(\frac{1}{ 2 \pi \sigma^2_{\zeta}} \right)^{-\frac{N-p}{2}} \exp \left\lbrace -\frac{1}{2 \sigma^2_{\zeta}} \sum_{t=p+1}^N \zeta^2_t(\Phi) \right\rbrace 
\end{equation}
where $\zeta_t(\Phi) = \beta_{k,t} - \sum_{i=1}^p \phi_i \beta_{k,t-i}$ and $\zeta_N = \left\langle \zeta_{p+1}, \cdots, \zeta_N \right\rbrace$. Let $\ell \left( \zeta_t(\Phi); \sigma^2_{\zeta} \right)  = \log \mathcal{L}  \left( \zeta_t(\Phi); \sigma^2_{\zeta} \right)$ denote the conditional log-likelihood function. Then, the maximum likelihood estimators of $\Phi$ and $\sigma^2_{\zeta}$ conditionally on the first $p$ observations, respectively, are obtained by the solution of the score functions $u_{\Phi}$ and $u_{\sigma_{\zeta}}$, as follows:
\begin{eqnarray*}
u_{\Phi} \left( \zeta_t(\Phi); \sigma^2_{\zeta} \right) &=& \frac{\partial}{\partial \Phi} \ell \left( \zeta_t(\Phi); \sigma^2_{\zeta} \right) \\
u_{\sigma_{\zeta}} \left( \zeta_t(\Phi); \sigma^2_{\zeta} \right) &=& \frac{\partial}{\partial \sigma_{\zeta}} \ell \left( \zeta_t(\Phi); \sigma^2_{\zeta} \right)
\end{eqnarray*}

The weighted likelihood methodology is proposed by \cite{Markatou1996} and \cite{Basu1998} to construct efficient and robust estimators by replacing the usual score functions with weighted score equations that measure the discrepancy between the estimated and hypothesized model densities. It has been extended to a wide variety of statistical inference problems, see, for example, \cite{Claudio2001}, \cite{Claudio2002a, Claudio2002b}, \cite{claudiobook} and \cite{Claudio2010}. Let us consider the probability density function given by (\ref{eq:dens}) and let $f^*\left( \zeta_t(\Phi), \hat{F}_N(\Phi) \right)$ and $m^*\left( \zeta_t(\Phi),\sigma^2_{\zeta} \right)$ denote a kernel density estimator with bandwidth $h$ based on the empirical distribution function $\hat{F}_N(\Phi)$ and the kernel smoothed model density, as follows:
\begin{eqnarray*}
f^*\left( \zeta_t(\Phi), \hat{F}_N(\Phi) \right) &=& \int k \left( \zeta_t(\Phi);r,g \right) d \hat{F}_N (r; \Phi) \\
m^*\left( \zeta_t(\Phi),\sigma^2_{\zeta} \right) &=& \int k \left( \zeta_t(\Phi);r,g \right) d M(r; \sigma^2_{\zeta})
\end{eqnarray*}
where $M(\sigma^2_{\zeta})$ and $k \left( \zeta_t(\Phi);r,g \right)$ represent the normal distribution function with mean zero and variance $\sigma^2_{\zeta}$ and a kernel density with bandwidth $g$, respectively. Denote by $\delta_t \left(\zeta_t(\Phi); M(\sigma^2_{\zeta}), \hat{F}_N(\Phi) \right) = \frac{f^*\left( \zeta_t(\Phi), \hat{F}_N(\Phi) \right)}{m^*\left( \zeta_t(\Phi),\sigma^2_{\zeta} \right)}$ and $\omega(\delta_t) = \min \left\lbrace 1, \frac{[A(\delta_t)+1]^{+}}{\delta_t + 1} \right\rbrace $ the Pearson residual and the weight function, respectively, where $[\cdot]^{+}$ denotes the positive part and $A(\cdot)$ is the residual adjustment function. Note that we consider the Hellinger residual adjustment function of \cite{Lindsay}, $A(\delta) = 2[(\delta+1)^{0.5}-1]$. Based on these definitions, the conditional weighted likelihood of the parameters $\Phi$ and $\sigma_{\zeta}$ are obtained by solving the following estimating equations:
\begin{eqnarray*}
& & (N-p)^{-1} \sum_{t=p+1}^N \omega(\delta_t) u_{\Phi} \left( \zeta_t(\Phi); \sigma^2_{\zeta} \right) \\
& & (N-p)^{-1} \sum_{t=p+1}^N \omega(\delta_t) u_{\sigma_{\zeta}} \left( \zeta_t(\Phi); \sigma^2_{\zeta} \right)
\end{eqnarray*} 

Now let $\hat{\Phi}^{\omega} = \left( \hat{\phi}^{\omega}_1, \cdots, \hat{\phi}^{\omega}_p \right)$ denote the estimated parameter vector of AR($p$) process using the weighted likelihood methodology. Denote by $\hat{\beta}^{\omega}_{k, N+h|N} = \sum_{i=1}^p \hat{\phi}^{\omega}_i \hat{\beta}^{\omega}_{k, N+h-i|N}$ the weighted likelihood version of $h$-step ahead forecast of $\beta_{k, N+h}$. Then, the weighted likelihood-based $h$-step-ahead point forecasts and bootstrap prediction intervals for $y_{N+h|N}$ are obtained similarly as in the algorithm given in the previous subsection. In the proposed forecasting strategy, the following holds for the final weights obtained from the full model, $\hat{\omega}_t$, when the model is accurately specified and no outlying observation is present in the data; $\sup_t \vert \hat{\omega}_t - 1 \vert \xrightarrow{p} 0$, see \cite{Claudio1998a}. Following by \cite{Claudio2002b}, it can be shown that $ \left| \sum_{t=1}^N \hat{\omega}_t \zeta_t(\hat{\Phi}_{\omega})^2 - \sum_{t=1}^N \zeta_t(\hat{\Phi})^2 \right| = o_p(N)$. This result indicates that 
\begin{inparaenum}
\item[($i$)] the weighted likelihood method tends to perform similar to the maximum likelihood method when no outlier is present in the data, and 
\item[($ii$)] the weighted likelihood is expected to have better performance than the maximum likelihood method when the data are contaminated by the outlier(s) (since it downweights the `bad' forecast errors caused by the contamination).
\end{inparaenum}

\section{Numerical results\label{sec:sim}}

\begin{figure}[htbp]
  \centering
  \includegraphics[width= \textwidth]{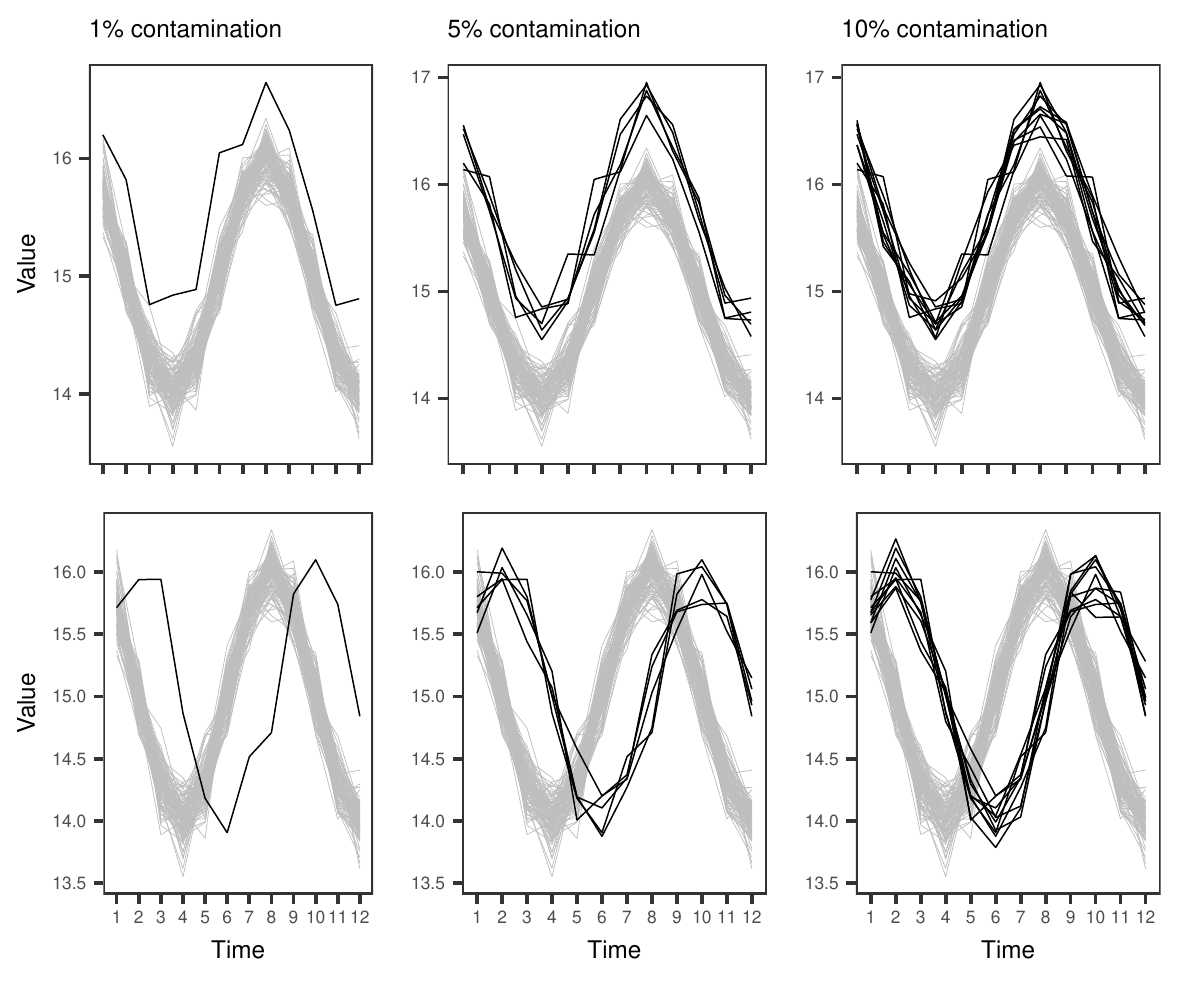}
  \caption{Example of the simulated functional time series (grey lines) with outliers (black lines): magnitude outlier (first row) and shape outlier (second row).}
  \label{fig:simdat}
\end{figure}

This section reports the finite sample performance of the proposed functional time series forecasting method via several Monte Carlo experiments. Throughout the experiments, two different simulation scenarios are considered: 
\begin{inparaenum}
\item[(1)] The data are generated from a relatively smooth process with no outlying observations, and 
\item[(2)] $\gamma = [1\%, 5\%, 10\%]$ of the generated data is contaminated by the deliberately inserted magnitude and shape outlier(s). The following process is used to generate the data; $y_t(x_j) = 15 + \cos(\pi j/4) + N(0, 0.15^2)$ ($t = 1, \cdots, N = 100$ and $j = 1, \cdots, J = 12$). The magnitude outlier(s) are generated by contaminating $N \gamma$ randomly selected function(s) by a random function $y^c_t(x_j) = \vert N(0.75, 0.15^2) \vert$, that is $y_t(x_j) = y_t(x_j) + y^c_t(x_j)$. Conversely, the shape outliers are generated from the following process: $15 + \sin(\pi j/4) + N(0, 0.15^2)$. Examples of the simulated data (with magnitude and shape outliers) are presented in Figure~\ref{fig:simdat}. 
\end{inparaenum}

We divide the generated data into two parts so that we can construct the model using the first $N-h$ functions to obtain point forecasts and the last $h$ functions for the bootstrap prediction intervals, where $h = [1, 5, 10]$. To construct an functional time series model, first, the noisy functional time series is converted to a smooth function by the smoothing spline method. Then, the hybrid principal component model of \cite{HydUl} is used to decompose the smooth functions. The nominal level $\alpha$ is set to 0.05 to calculate 95\% bootstrap prediction intervals. For each scenario, $MC = 1,000$ Monte Carlo simulations with $B = 999$ bootstrap resamples are performed, and only the first $K = 3$ principal components and their scores are used to obtain forecasts. The performance of the proposed method is compared with that of the autoregressive integrated moving average (ARIMA) model as used by \cite{HydUl} and most of the references cited. To evaluate the forecasting performance of the proposed method, we calculate several performance metrics, including the average mean squared error (AMSE), coverage probability (Cp) and the average interval scores (Score), as follows:
\begin{eqnarray}
\text{AMSE} &=& \frac{1}{h \times 12} \sum_{i=1}^h \sum_{j=1}^{12} \left\lbrace  y_{N+h}(x_j) - \hat{y}_{N+h|N}(x_j) \right\rbrace^2 \nonumber \\
\text{Cp} &=& \frac{1}{h \times 12} \sum_{i=1}^h \sum_{j=1}^{12} 1 \left\lbrace \hat{y}^{*, 0.025}_{N+h|N}(x_j) \leq y_{N+h}(x_j) \leq \hat{y}^{*, 0.975}_{N+h|N}(x_j) \right\rbrace  \nonumber \\
\text{Score} &=& \frac{1}{h \times 12} \sum_{i=1}^h \sum_{j=1}^{12} \left\lbrace \left( \hat{y}^{*, 0.975}_{N+h|N}(x_j) - \hat{y}^{*, 0.025}_{N+h|N}(x_j) \right) \right. \nonumber \\
&+& \frac{2}{\alpha} \left( \hat{y}^{*, 0.025}_{N+h|N}(x_j) - y_{N+h}(x_j) \right) \mathds{1} \left\lbrace y_{N+h}(x_j) < \hat{y}^{*, 0.025}_{N+h|N}(x_j) \right\rbrace \nonumber \\
&+& \left. \frac{2}{\alpha} \left( y_{N+h}(x_j) - \hat{y}^{*, 0.975}_{N+h|N}(x_j)\right) \mathds{1} \left\lbrace y_{N+h}(x_j) > \hat{y}^{*, 0.975}_{N+h|N}(x_j)\right\rbrace \nonumber \right\rbrace 
\end{eqnarray}
where $\mathds{1} \left\lbrace \cdot \right\rbrace$ denotes the indicator function and $\hat{y}^{*, \alpha}_{N+h|N}(x_j)$ is the $\alpha$\textsuperscript{th} quantile of the bootstrap replicates.

The results obtained from the Monte Carlo experiments are reported in Table~\ref{tab:sim}. Our findings show that, regardless of the forecast horizon, both weighted likelihood and ARIMA-based forecasting models tend to have similar AMSE values for the point forecasts when no outliers are present in the data. However, the proposed method produces better coverage probabilities and interval scores than those obtained from ARIMA. These results indicate that compared with ARIMA, we can obtain more reliable and more precise prediction intervals using the weighted likelihood-based bootstrap method even if the outliers do not contaminate the data. When the data has outlier(s) (magnitude and/or shape), both methods still have similar AMSE values for the point forecasts when $h=1$ but the proposed method has better coverage probabilities and interval scores compared with those of ARIMA. By contrast, for long-term forecast horizons ($h=5$ and $h=10$), the proposed method has smaller AMSE values than ARIMA, and the difference becomes more prominent as the forecast horizon increases. For the prediction intervals, while it seems that both methods have similar coverage performance, the weighted likelihood-based bootstrap method produces significantly narrower prediction intervals than those of ARIMA. This is because the bootstrap method based on ARIMA is considerably affected by the large forecast errors produced by the outliers. However, the proposed method downweights the effects of outliers, and the structure of the bootstrap prediction intervals are not distorted. 

\begin{center}
\tabcolsep 0.245in
\begin{longtable}{@{}llllccc@{}}
\caption{Simulation results: Estimated AMSE, Cp and Score values for the proposed (weighted likelihood estimator (WLE)) and ARIMA models when the data have magnitude (MO) and shape (SO) outliers.}\label{tab:sim}\\
\toprule
{Contamination} & $h$ & {Outlier} & {Method} & {AMSE} & {Cp} & {Score} \\
\endfirsthead
{Contamination} & $h$ & {Outlier} & {Method} & {AMSE} & {Cp} & {Score} \\
\toprule
\endhead
\multicolumn{7}{r}{{Continued on next page}}\\
\endfoot
\endlastfoot
\midrule
\multirow{12}{*}{0\%} & \multirow{2}{*}{$h=1$} & & WLE & 0.0228 & 0.9365 & 0.7244 \\
& & & ARIMA & 0.0229 & 0.9119 & 0.7419 \\ 
\cmidrule{4-7}
& \multirow{2}{*}{$h=5$} & & WLE & 0.0228 & 0.9360 & 0.7258 \\
& & & ARIMA & 0.0228 & 0.9122 & 0.7413 
\\
\cmidrule{4-7}
& \multirow{2}{*}{$h=10$} & & WLE & 0.0229 & 0.9350 & 0.7312 \\
& & & ARIMA & 0.0228 & 0.9106 & 0.7492 \\ 
\hline
\multirow{12}{*}{1\%} & \multirow{4}{*}{$h=1$} & MO & WLE & 0.0229 & 0.9375 & 0.7299 \\
& & & ARIMA & 0.0229 & 0.9245 & 0.7507 \\ \cmidrule{3-7}
& & SO & WLE & 0.0238 & 0.9153 & 0.7525 \\
& & & ARIMA & 0.0242 & 0.9086 & 0.7749 \\
\cmidrule{2-7}
& \multirow{4}{*}{$h=5$} & MO & WLE & 0.0231 & 0.9438 & 0.7303 \\
& & & ARIMA & 0.0246 & 0.9553 & 1.1374 \\ \cmidrule{3-7}
& & SO & WLE & 0.0225 & 0.9366 & 0.7315 \\
& & & ARIMA & 0.0258 & 0.9413 & 1.3039 \\
\cmidrule{2-7}
& \multirow{4}{*}{$h=10$} & MO & WLE & 0.0242 & 0.9467 & 0.7404 \\
& & & ARIMA & 0.0296 & 0.9594 & 1.2357 \\ \cmidrule{3-7}
& & SO & WLE & 0.0228 & 0.9373 & 0.7398 \\
& & & ARIMA & 0.0343 & 0.935 & 1.4226 \\
\hline
\multirow{12}{*}{5\%} & \multirow{4}{*}{$h=1$} & MO & WLE & 0.0229 & 0.9378 & 0.7277 \\
& & & ARIMA & 0.0230 & 0.9246 & 0.7415 \\ \cmidrule{3-7}
& & SO & WLE & 0.0233 & 0.9167 & 0.7537 \\
& & & ARIMA & 0.0242 & 0.9089 & 0.7847 \\
\cmidrule{2-7}
& \multirow{4}{*}{$h=5$} & MO & WLE & 0.0233 & 0.9420 & 0.7329 \\
& & & ARIMA & 0.0249 & 0.9557 & 1.1344 \\ \cmidrule{3-7}
& & SO & WLE & 0.0297 &0.9306 & 0.7375 \\
& & & ARIMA & 0.0263 & 0.9342 & 1.3158 \\
\cmidrule{2-7}
& \multirow{4}{*}{$h=10$} & MO & WLE & 0.0233 & 0.9507 & 0.7298 \\
& & & ARIMA & 0.0295 & 0.9592 & 1.2390 \\ \cmidrule{3-7}
& & SO & WLE & 0.0247 & 0.9346 & 0.7583 \\
& & & ARIMA & 0.0361 & 0.9322 & 1.4349 \\
\hline
\multirow{12}{*}{10\%} & \multirow{4}{*}{$h=1$} & MO & WLE & 0.0227 & 0.9377 & 0.7265 \\
& & & ARIMA & 0.0228 & 0.9244 & 0.7476 \\ \cmidrule{3-7}
& & SO & WLE & 0.0228 & 0.9197 & 0.7601 \\
& & & ARIMA & 0.0238 & 0.9071 & 0.8843 \\
\cmidrule{2-7}
& \multirow{4}{*}{$h=5$} & MO & WLE & 0.0232 & 0.9437 & 0.7315 \\
& & & ARIMA & 0.0249 & 0.9546 & 1.1469 \\ \cmidrule{3-7}
& & SO & WLE & 0.0231 & 0.9225 & 0.7652 \\
& & & ARIMA & 0.0276 & 0.9217 & 1.2838 \\
\cmidrule{2-7}
& \multirow{4}{*}{$h=10$} & MO & WLE & 0.0236 & 0.9491 & 0.7324 \\
& & & ARIMA & 0.0303 & 0.9582 & 1.2457 \\ \cmidrule{3-7}
& & SO & WLE & 0.0232 & 0.9286 & 0.7769 \\
& & & ARIMA & 0.0374 & 0.9338 & 1.4522 \\
\bottomrule
\end{longtable}
\end{center}

The finite-sample performance of a maximum likelihood based forecasting method, such as ARIMA, depends on the magnitude of outliers. Via several Monte-Carlo experiments, we examine the effects of outlier size on forecasting accuracy of the ARIMA and weighted likelihood based functional time series methods. Our results indicate that the performance of the ARIMA becomes worse as the magnitude of the outliers increases while the proposed weighted likelihood based functional time series method produces consistent results. For instance, when the magnitude outliers are generated from $y^c_t(x_j) = \vert N(3.75, 0.15^2) \vert$, the one-step-ahead forecasting performances of the ARIMA and proposed weighted likelihood based functional time series methods are presented in Table~\ref{tab:sim2}. Compared to the results reported in Table~\ref{tab:sim}, Table~\ref{tab:sim2} shows that when $[1\%, 5\%, 10\%]$ of the data are contaminated with larger magnitude outliers, the ARIMA method produced about $[1.40, 2.76, 7.91]$ times larger AMSE than the ones obtained when the data are contaminated with small magnitude outliers and $[1.06, 5.64, 5.96]$ times larger score values. On the other hand, the proposed method produced almost the similar AMSE and score values in both cases. Similar results can be obtained from the corresponding author upon request for the case when the data have large shape outliers.

\begin{center}
\tabcolsep 0.45in
\begin{longtable}{@{}llllccc@{}}
\caption{Simulation results: Estimated AMSE, Cp and Score values for the proposed (weighted likelihood estimator (WLE)) and ARIMA models when the data have large magnitude outliers and forecast horizon $h = 1$.}\label{tab:sim2}\\
\toprule
{Contamination} & {Method} & {AMSE} & {Cp} & {Score} \\
\endfirsthead
{Contamination} & {Method} & {AMSE} & {Cp} & {Score} \\
\toprule
\endhead
\multicolumn{7}{r}{{Continued on next page}}\\
\endfoot
\endlastfoot
\midrule
\multirow{2}{*}{1\%} & WLE & 0.0300 & 0.9352 & 0.7575 \\
& ARIMA & 0.0321 & 0.9100 & 0.8003 \\
\hline
\multirow{2}{*}{5\%} & WLE & 0.0229 & 0.9533 & 0.7564 \\
& ARIMA & 0.0636 & 0.9710 & 4.1816 \\
\hline
\multirow{2}{*}{10\%} & WLE & 0.0305 & 0.9612 & 0.9618 \\
& ARIMA & 0.1805 & 0.9780 & 4.4548 \\
\bottomrule
\end{longtable}
\end{center}

\section{Real-data examples\label{sec:real}}

\begin{figure}[htbp]
  \centering
  \includegraphics[width= \textwidth]{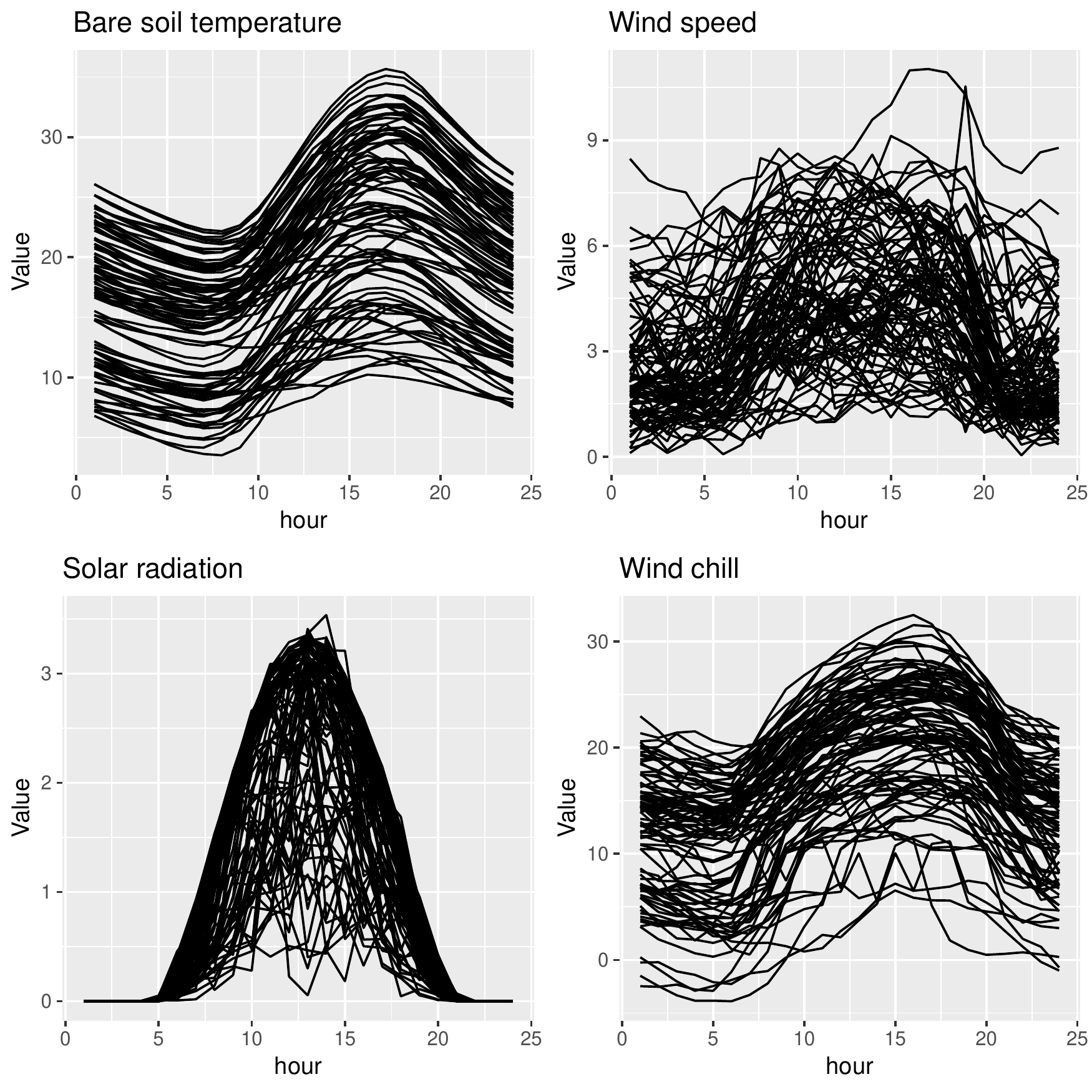}
  \caption{Functional time series representations of the environmental variables.}
  \label{fig:real-fts}
\end{figure}

This section evaluates the finite sample performance of the proposed method using four environmental datasets---hourly bare soil temperature, wind speed, solar radiation, and wind chill from 1 May 2017 to 31 July 2017 (92 days in total)---which are collected from the Michigan weather station (the data are obtained from North Dakota Agricultural Weather Network Center  \url{https://ndawn.ndsu.nodak.edu/}). The functional time series representation of the datasets is presented in Figure~\ref{fig:real-fts}. It is clear from this figure that the bare soil temperature functions are reasonably smooth and have no clear outlier. Conversely, the functions of the functional time series of wind speed, solar radiation, and wind chill are noisy, and all three datasets have several outlying functions. For all four datasets, we obtain only one-step-ahead forecasts. The forecasting performance of the ARIMA and of the proposed method are compared using the rolling holdout testing samples, in line with \cite{Shang2019R} and \cite{HydUl}; thus, 80\% of the datasets are used as training samples and the remaining 20\% are used for validation. In the modelling step, the number of principal components is also determined based on rolling holdout validation samples. The optimal value of $K$ is found as 4 for the bare soil temperature dataset and 6 for the other three datasets.

\begin{figure}[htbp]
\begin{center}
\begin{tabular}{cc}
  \includegraphics[width=80mm]{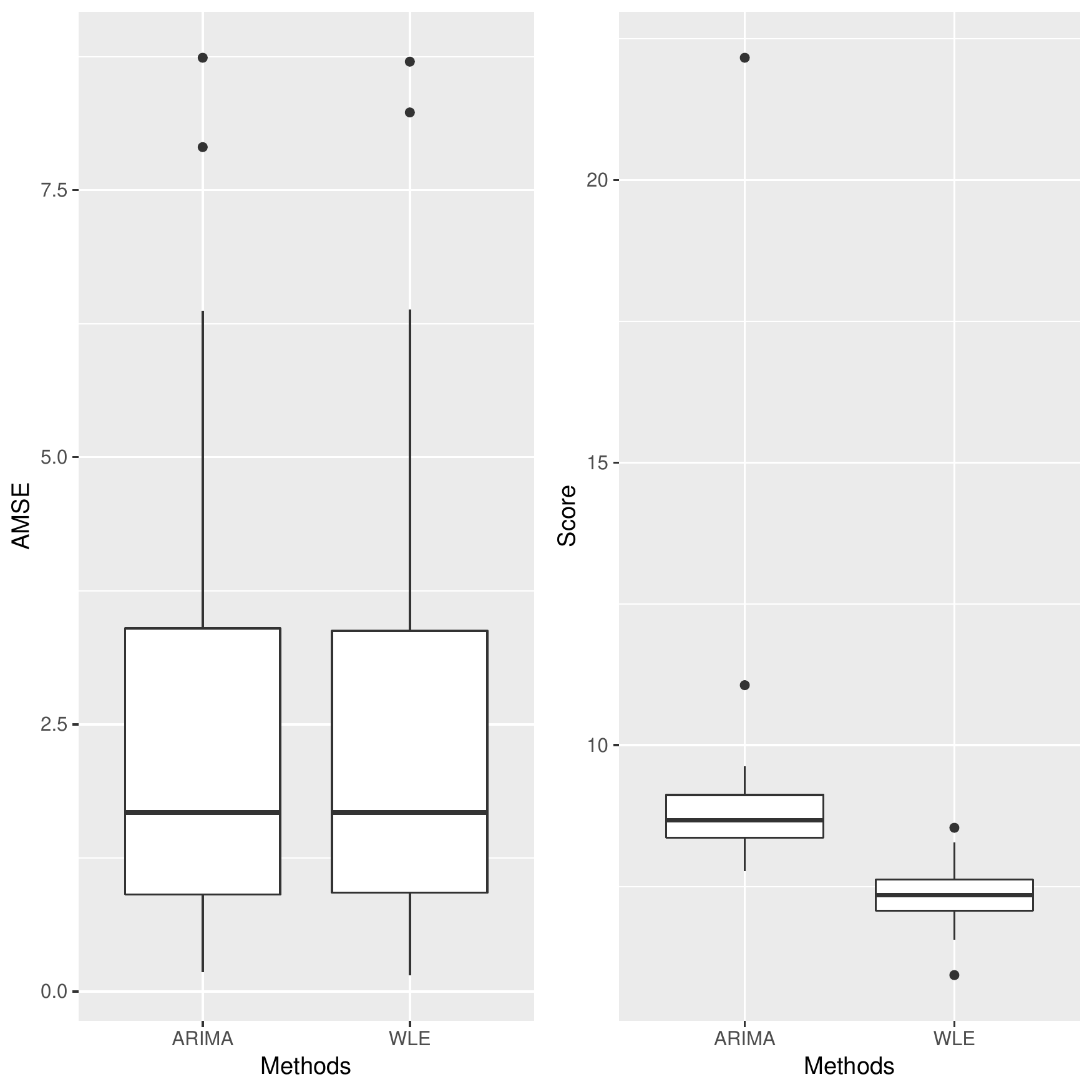} &   \includegraphics[width=80mm]{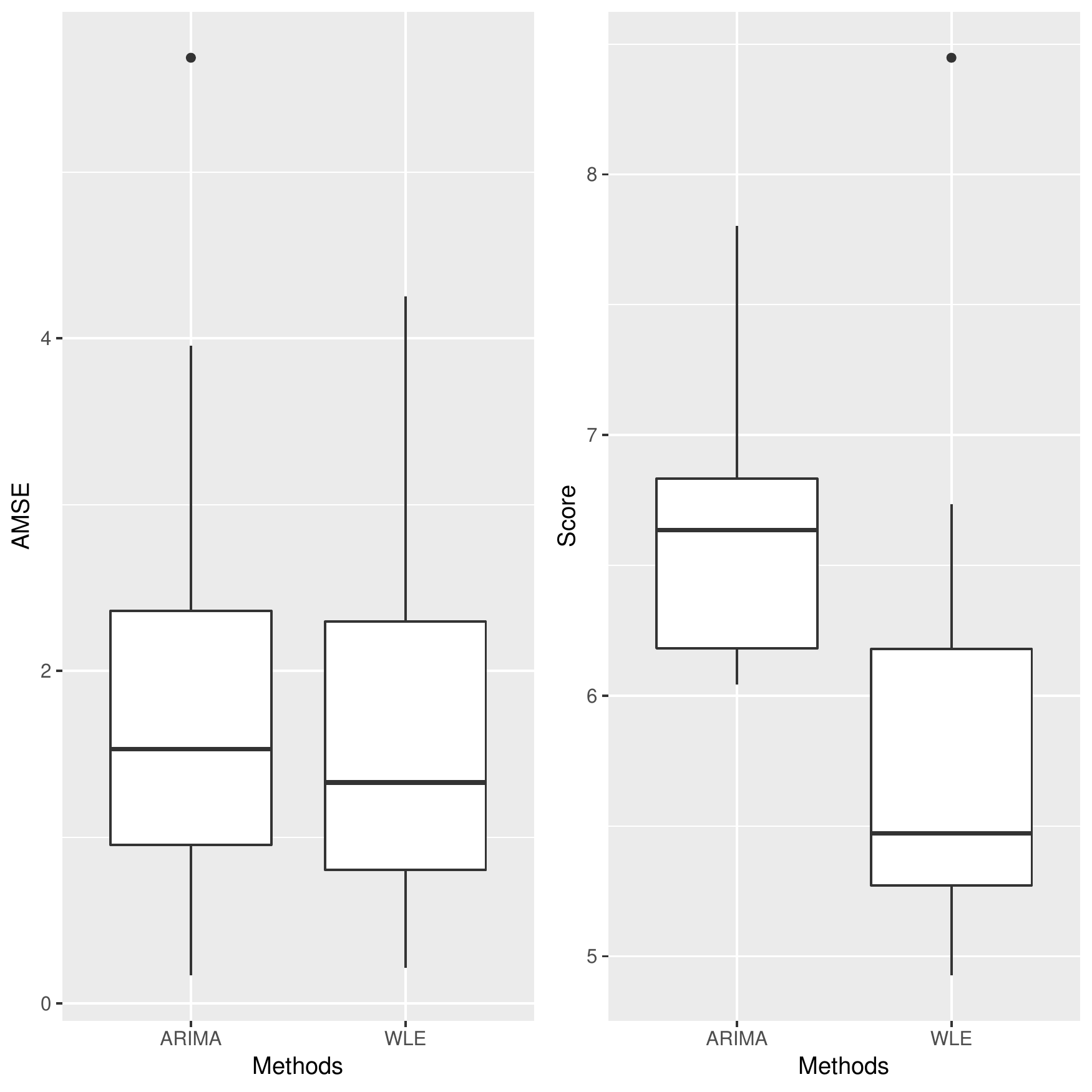} \\
 Bare soil temperature &  Wind speed \\[6pt]
 \includegraphics[width=80mm]{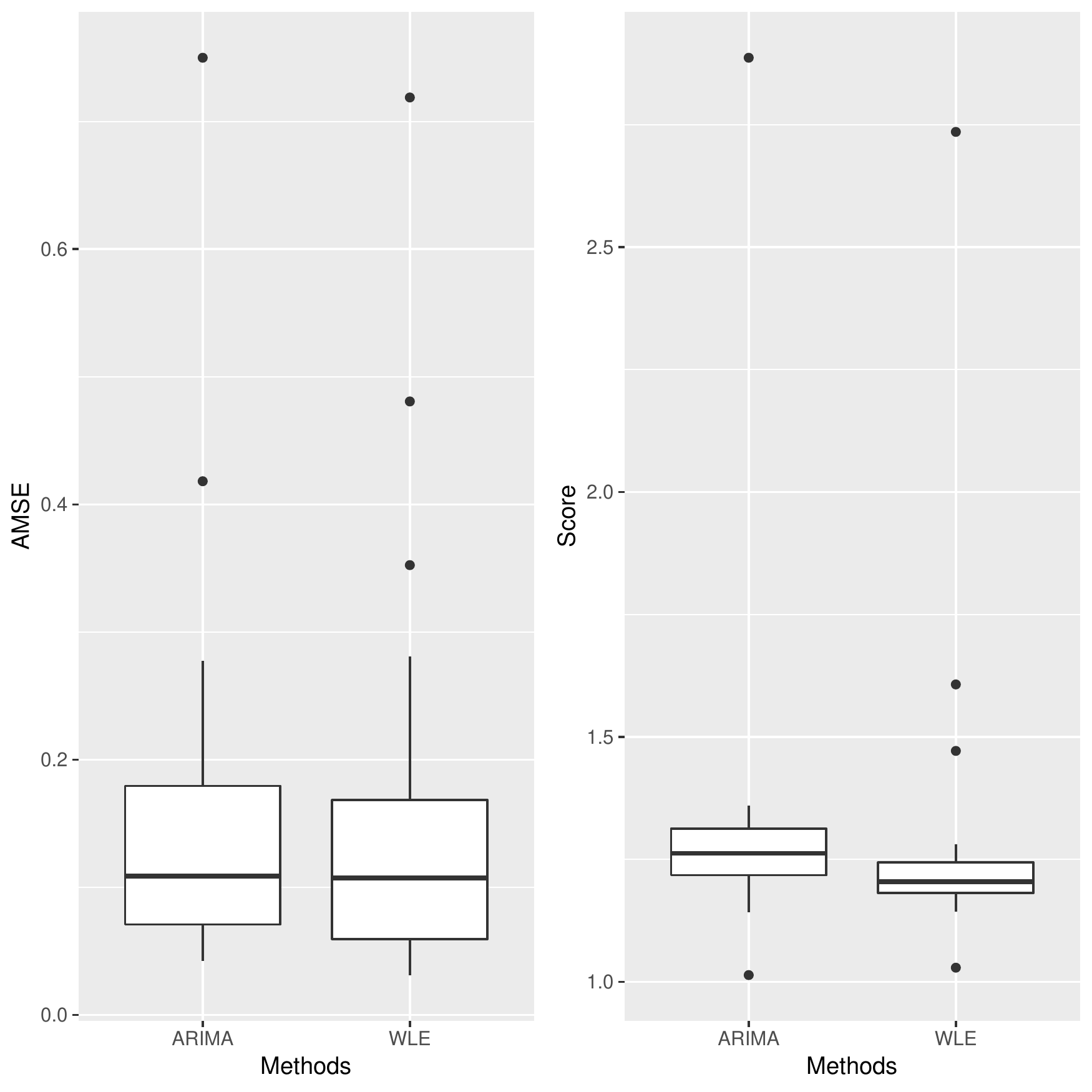} &   \includegraphics[width=80mm]{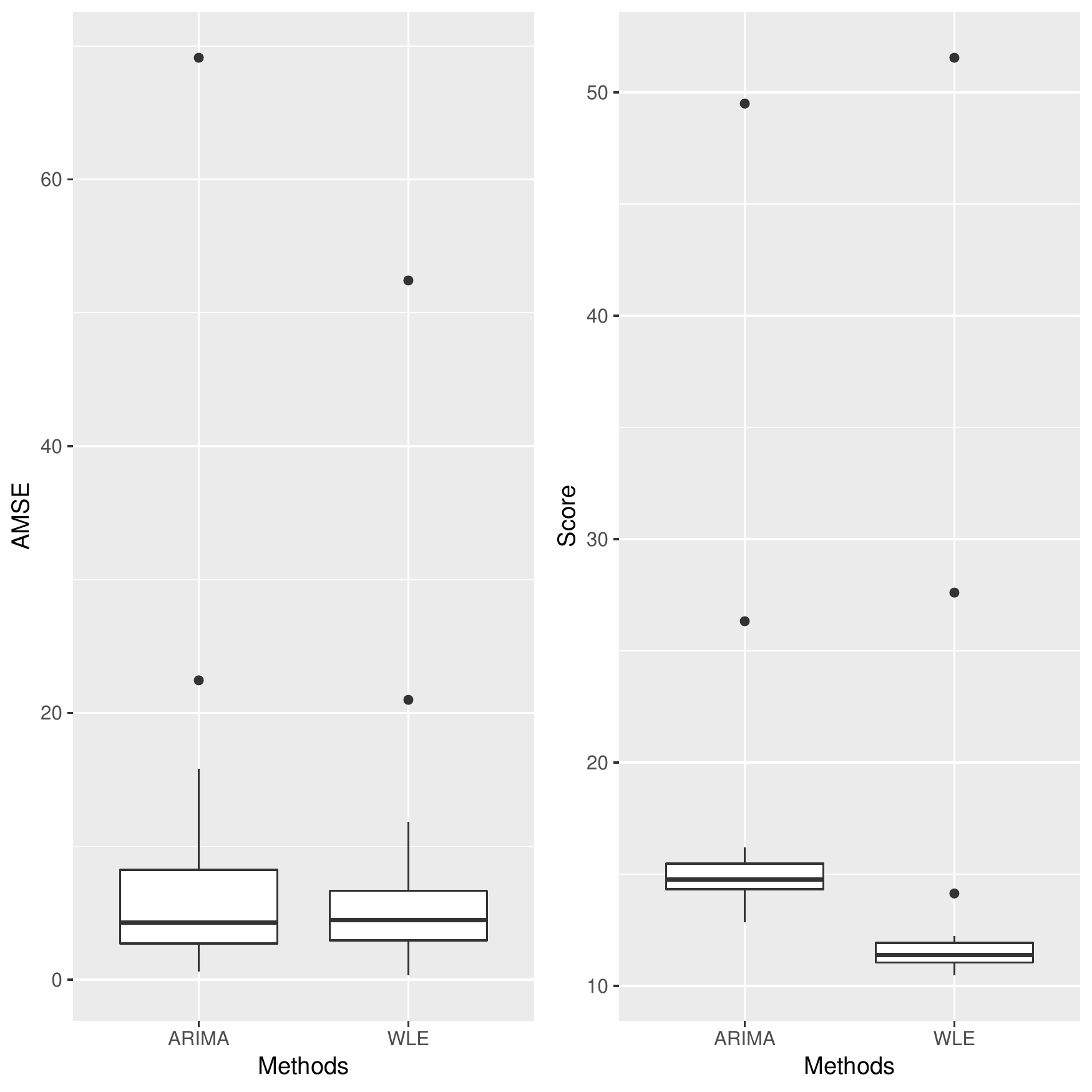} \\
 Solar radiation &  Wind chill \\[6pt]
\end{tabular}
\caption{Obtained AMSE and Score values of the ARIMA and proposed (WLE) models for the real-data examples.}
\label{fig:boxplots}
\end{center}
\end{figure}

\begin{figure}[htbp]
  \centering
  \includegraphics[width= \textwidth]{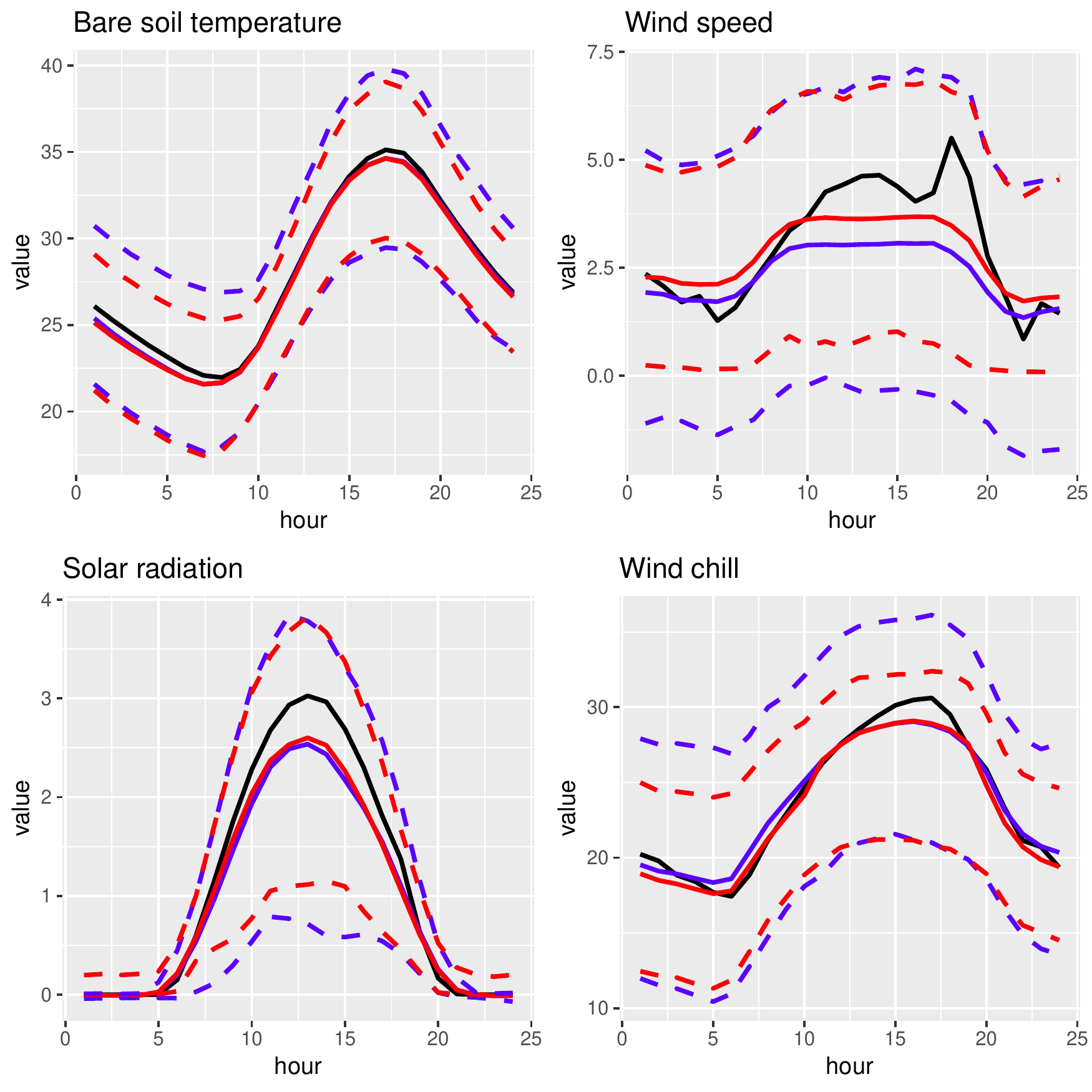}
  \caption{Obtained one-step-ahead point forecasts: ARIMA (blue solid lines) and weighted likelihood (red solid lines), and bootstrap 95\% prediction intervals: ARIMA (blue dashed lines) and weighted likelihood (red solid lines), together with the observed function (black solid lines).}
  \label{fig:real-frcst}
\end{figure}

The obtained AMSE and Score values are presented in Figure~\ref{fig:boxplots}. The results indicate that compared with ARIMA, the proposed method produces slightly better AMSE values for the one-step-ahead point forecast. The results also show that the weighted likelihood-based bootstrap method produces better prediction intervals than those obtained from the ARIMA-based bootstrap procedure especially for the datasets contaminated by outliers. Figure~\ref{fig:real-frcst} presents a graphical representation of the one-step-ahead forecasts for which the functional time series models are constructed based on the first 91 functions. This figure supports the results presented in Figure~\ref{fig:boxplots}. 

\section{Conclusion\label{sec:conc}}

functional time series are frequently observed in many scientific fields owing to novel data collection tools. Consequently, several functional time series models have been developed to analyse such datasets and to obtain forecasts of their unobserved realizations. Recent studies have shown that the univariate time series models together with functional principal component regression can be used to obtain valid point forecasts and bootstrap prediction intervals. However, the traditional univariate time series models such as ARIMA, which are commonly used in the analyses, may severely be affected in the presence of outliers, leading to poor forecasting results. Therefore, we propose a robust functional time series forecasting approach based on the weighted likelihood methodology. We evaluate the finite sample performance of the proposed method using several Monte Carlo experiments and four environmental datasets. The numerical results produced showed that the proposed method is a good competitor for ARIMA and would be widely adopted because of its narrower prediction intervals.

\newpage
\bibliographystyle{tandfx}
\bibliography{mybibfile}

\end{document}